\documentclass[aps,prd,preprint,showpacs]{revtex4}

\usepackage{graphics}
\usepackage{epsfig}
\usepackage{latexsym}
\usepackage{colordvi}
\usepackage{amsmath}
\usepackage{amssymb}

\begin{document}


\draft

\title{The Roper and Radiative Decay of Pentaquarks}

\author{Deog Ki Hong}
\email[E-mail: ]{dkhong@pusan.ac.kr,   dkhong@lns.mit.edu}
\affiliation{Department of Physics, Pusan National University,
             Busan 609-735, Korea}
\affiliation{Center for Theoretical Physics, Massachusetts
             Institute of Technology, Cambridge, MA 02139, USA}

\vspace{0.1in}

\date{\today}

\begin{abstract}
Identifying the Roper ${N}(1440)$ and ${N}(1710)$ as
the pentaquark octet and anti-decuplet, respectively, we
analyze their main decay modes in the diquark picture. The ratio of
the partial decay widths
is largely consistent with the nearly ideal mixing of
the Jaffe-Wilczek diquark model, 
which then allows to predict  the width of the radiative decay of ${N}(1440)$,
$\Gamma_{10}({N}\to p\gamma)=1/4\,
\Gamma_{12}({N}\to n\gamma)=0.25-0.31~{\rm MeV}$.
We then show that
the three-body radiative decay of the pentaquark anti-decuplet is
quite enhanced due to its mixing with the pentaquark octet.
We find for the $J^P={\frac{1}{2}}^+$ pentaquark anti-decuplet
$\Gamma(\Theta^+\to K^+\,n\,\gamma)=0.034\sim0.041~{\rm MeV}$.
The diquark picture of the pentaquark predicts
$\Gamma(\Theta^+\to K^+\,n\,\gamma) =4\,\Gamma(\Theta^+\to K^0\,p\,\gamma)$.
Finally we show that the difference in the $\Theta^+$ mass in the $K^0p$ and
$K^+n$ decay channels
may be accounted for by the missing photons in the radiative decay.
\end{abstract}


\maketitle


Since the discovery of a narrow state of exotic baryons
in the photon-nucleon scattering at LEPS~\cite{Nakano:2003qx}
as predicted by chiral soliton models~\cite{Diakonov:1997mm},
several models have been proposed for exotic baryons~\cite{Jaffe:2004ph}.
However,
in spite of the active theoretical study on exotic baryons,
the current experimental situation of pentaquarks~\cite{trilling}
is quite confusing, as some of the
subsequent  experiments have not seen them.

In this letter, we analyze the decay modes of ${N}(1440)$ and
${N}(1710)$ to show that they
are consistent with the Jaffe-Wilczek (JW) diquark model~\cite{Jaffe:2003sg}.
We then point out that the three-body radiative decay of the pentaquark
anti-decuplet can be quite enhanced due to the near degeneracy between
the octet and the anti-decuplet and also the diquark nature of pentaquarks.
Since  the radiative decay amplitude is proportional
to the electric charge of the anti-quark bound in the pentaquark,
the diquark model predicts $\Gamma(\Theta^+\to\, K^+\,n\,\gamma)
=4\,\Gamma(\Theta^+\to\, K^0\,p\,\gamma)$, which will be a clear signal
for the diquark models. 



In the JW model, the exotic baryons are
bound states of two scalar diquarks and one anti-quark, forming
the multiplets of low dimensions, $\overline{10}$ and $8$, of the SU(3) flavor symmetry.
The degeneracy of $\overline{10}$ and $8$ is lifted by a nearly
ideal mixing between them. It was soon confirmed by
NA49 experiments~\cite{Alt:2003vb}, which discovered $S=-2$ pentaquarks
of mass  $1860~{\rm MeV}$, not much different from the JW prediction.
The decay widths of pentaquarks are also explicitly calculated in the
diquark effective theory based on the JW model to
find they are naturally small,
a few MeV or less, since they decay through tunnelling
the potential barrier between two diquarks~\cite{Hong:2004xn}.

In the JW model, the lightest member of the
pentaquark octet is identified, after mixing,
as the Roper state $N(1440)$. Its orthogonal state
in the anti-decuplet is identified as $N(1710)$.
Though the assignment fits well in the mass formula,
it is argued in~\cite{Cohen:2004gu,Goeke:2004ht} that such an identification
leads to a gross violation of SU(3) in partial decay widths,
unseen in other hadronic decays.
However, we show that the partial decay widths of ${ N}(1440)$
and ${N}(1710)$ are consistent with the JW model, because
the gross violation is due to the fact that the exotic baryons
decay by tunneling, whose amplitude  depends exponentially on the SU(3) violating terms.
Another concern~\cite{Mohta:2004xg}
was that the decay branching fraction for $\Delta(1600)\to{ N}(1440)\,\pi$
is not suppressed at all, $\Gamma_1/\Gamma=10-25\%$~\cite{Eidelman:2004wy}.
But, since the exotic baryon production is suppressed not by tunneling
but only by the production amplitude of diquarks,
whose strength is nothing but the Yukawa coupling
of the diquark to two quarks, $g\approx1.7$~\cite{Hong:2004xn},
the observed branching ratio does not necessarily contradict
with the diquark picture.

Finally, we show that a three-body radiative decay is quite
enhanced and contributes significantly to the decay of
the pentaquark anti-decuplet. 
In the JW model, due to the
mixing, the diquarks inside the octet are closer to each other
than in the anti-decuplet, making tunnelling easier for the octet.
Furthermore, since the octet and the anti-decuplet are almost
degenerate, the virtual octet in the three-body decay of the
anti-decuplet is near on-shell and makes the three-body decay more efficient.


In the ideal mixing, the quark content of the $S=0$ component of
the pentaquark octet is ${\cal N}_1=\left|\bar
q,\varphi_{ud},\varphi_{ud}\right>$, where $\bar q$ is the light
anti-quark and $\varphi_{ud}$ is the diquark made of $u$ and $d$,
while the $S=0$ component of the pentaquark anti-decuplet is
${\cal N}_{2}=\left|\bar s,\varphi_{qs},\varphi_{ud}\right>$,
where the $\varphi_{qs}$ diquark is made of the light quark $q$
and the strange quark, $s$.

The dominant decay modes of ${\cal N}_1$ and ${\cal N}_{2}$
 are shown in the diquark picture in
Fig.~{\ref{fig1_1}.
\begin{figure}[h]
\vskip 0.1in
{\centerline{\epsfxsize=2in\epsffile{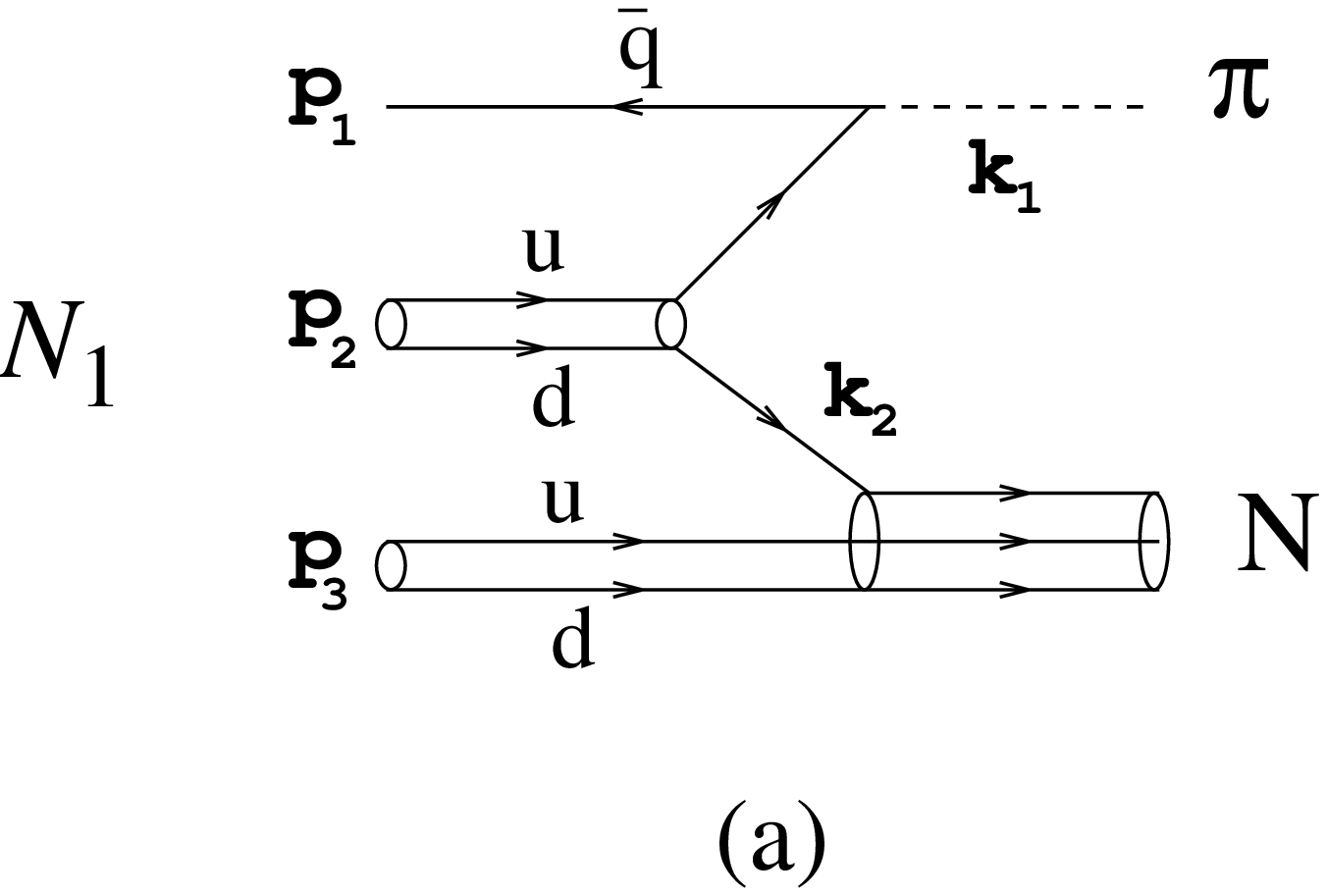}
\hskip 1in {\epsfxsize=2in \epsffile{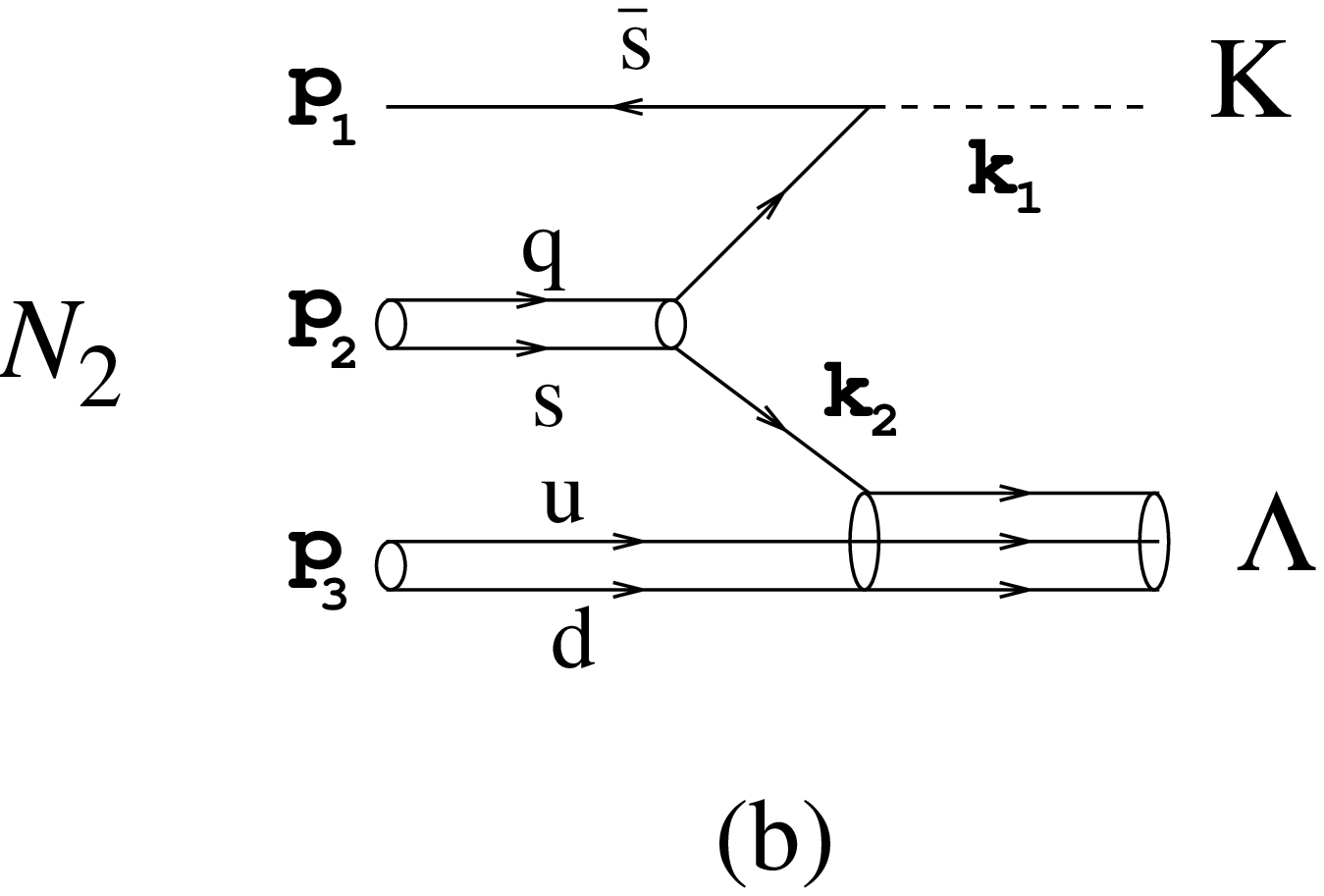}}
}}
\caption{The dominant decay modes of excited nucleons: (a) ${\cal N}_1\to\,N\pi$
(b) ${\cal N}_{2}\to\,\Lambda\,K$.
The breaking of diquark occurs through tunneling.}
 \label{fig1_1}
\end{figure}
From the viewpoint of the chiral perturbation theory, the processes shown in
Fig.~{\ref{fig1_1}}~(a) and (b) should not differ by more than 30\%,
the typical size of SU(3) breaking terms. 
However, the experimental data
differs by an order of magnitude, if ${\cal N}_1$ and
${\cal N}_{2}$ are identified as $N(1440)$ and $N(1710)$,
respectively: $\Gamma_a\left(N(1440)\to\,N\,\pi\right)=210-245~{\rm MeV}$
and $\Gamma_b\left(N(1710)\to\,\Lambda\,K\right)=5-25~{\rm MeV}$.
This may seem to contradict with the diquark picture. But,
a gross difference in the decay widths of two processes is expected,
since the SU(3) violating terms appear in the exponent of the tunneling
amplitude in the diquark picture.

The decay widths are given respectively as
\begin{eqnarray}
\Gamma_a\!\!&=&\!\!\frac{g_{\pi N{\cal N}_{1}}^2}
{4\pi\,f_{\pi}^2}\frac{|\vec k_1|\left(M_{N(1440)}+m_N\right)^2}{M_{N(1440)}^2}
\left[\left(M_{N(1440)}-m_N\right)^2-m_{\pi}^2\right]
\\
\Gamma_b\!\!&=&\!\!\frac{g_{K\Lambda{\cal N}_{2}}^2}
{4\pi\,f_{K}^2}\frac{|\vec k_1^{\prime}|
\left(M_{N(1710)}+m_{\Lambda}\right)^2}
{M_{N(1710)}^2}
\left[\left(M_{N(1710)}-m_{\Lambda}\right)^2-m_{K}^2\right],
\end{eqnarray}
where the outgoing meson momenta are $|\vec k_1|=396~{\rm MeV}$ and
$|\vec k_1^{\prime}|=268~{\rm MeV}$, respectively.
From the partial widths by the Particle Data Group (PDG),
we get $g_{\pi N{\cal N}_1}=\pm\,(0.3\sim0.33)$ and
$g_{K \Lambda{\cal N}_{2}}=\pm\,(0.10\sim0.23)$.

The amplitude for exotic baryons to decay into normal baryons is
quite suppressed due to the tunneling barrier for
the diquarks~\cite{Hong:2004xn}. Therefore, the couplings in the decay
of ${\cal N}_1$ and ${\cal N}_{2}$
have to be proportional to their tunneling amplitudes.
The WKB approximation for the tunneling amplitude for ${\cal N}_1$
gives
$e^{-S_0({\cal N}_1)}\simeq e^{-\Delta\,E\,r_0}=0.26$,
where the potential barrier is approximately the binding
energy of quarks in the scalar diquark, $\Delta E\simeq 270~{\rm MeV}$,
and $r_0\simeq(201~{\rm MeV})^{-1}$ is the average distance between
two diquarks of ${\cal N}_1$ in $P$-wave. The distance is estimated
from the mass formula for the naive diquark model
\begin{equation}
M_{{\cal N}_1}=2\,M_{ud}+m_q+\frac{1}{M_{ud}\,r_0^2},
\end{equation}
where $M_{ud}\simeq450~{\rm MeV}$ is the diquark $\varphi_{ud}$
mass~\cite{Hong:2004xn} and $m_q=360~{\rm MeV}$ is the constituent
mass of light quarks.
Similarly, the average distance between two diquarks $\varphi_{qs}$
and $\varphi_{ud}$ of ${\cal N}_{2}$ in $P$-wave
is estimated to be $r_1\simeq(145~{\rm MeV})^{-1}$ and
the ${\cal N}_{2}$ tunneling amplitude
$e^{-S_0({\cal N}_{2})}=0.16$.
Indeed, we find that the ratio of the tunneling amplitudes
is close to the ratio of the couplings,
\begin{equation}
\frac{e^{-S_0({\cal N}_1)}}{e^{-S_0({\cal N}_{2})}}
\simeq1.7,\quad \left|\frac{g_{\pi N{\cal N}_1}}
{g_{K \Lambda{\cal N}_{2}}}\right|\simeq 1.3-3.3
\end{equation}

It seems that most partial decay modes of ${N}(1710)$ are largely
consistent with the JW model. However, one of the dominant decay modes,
${N}(1710)\to\,N\pi$, does not fit in the JW model. The decay width
is given in the chiral perturbation theory as
$\Gamma({N}(1710)\to\,N\pi)=g_{\pi N{\cal N}_{2}}^2 7422~{\rm MeV}$,
where $g_{\pi N{\cal N}_{2}}$ is the coupling of
${\cal N}_{2}$ to $\pi\,N$.
In the ideal mixing,
${\cal N}_{2}=\left|\bar s,\varphi_{sq},\varphi_{ud}\right>$
does not decay into $N\pi$ at the leading
order~\cite{Hong:2004xn,Hong:2004ux}.
However, the decay width is estimated to be $10-20~{\rm MeV}$
by PDG. This can be
resolved if the mixing between the anti-decuplet and the octet is not exactly ideal
but nearly ideal~\cite{Mohta:2004xg}; $\theta=\cos^{-1}\sqrt{2/3}+\delta$.
Then
${\cal N}_{2}
\simeq
\left|\bar s,\varphi_{sq},\varphi_{ud}\right>
+\delta\left|\bar q,\varphi_{ud},\varphi_{ud}\right>$ 
and the partial decay width becomes
\begin{equation}
\Gamma({N}(1710)\to\,N\pi)
\approx\delta^2\,\Gamma({N}(1440)\to\,N\pi),
\end{equation}
which gives
$\delta=\pm\left|g_{\pi N{\cal N}_{2}}/g_{\pi N{\cal N}_1}\right|\approx\pm\, 0.26$
or the mixing angle $\theta\approx 20^{\circ}$.



We first examine the radiative decay of the Roper ${N}(1440)$.
Assuming it is the mixed pentaquark  ${\cal N}_1$, we have drawn
the diagram for the radiative
decay in Fig.~\ref{fig2}~(a), neglecting the angle $\delta$.
\begin{figure}[h]
\vskip 0.1in
\epsfxsize=2.2in
{\centerline{\epsffile{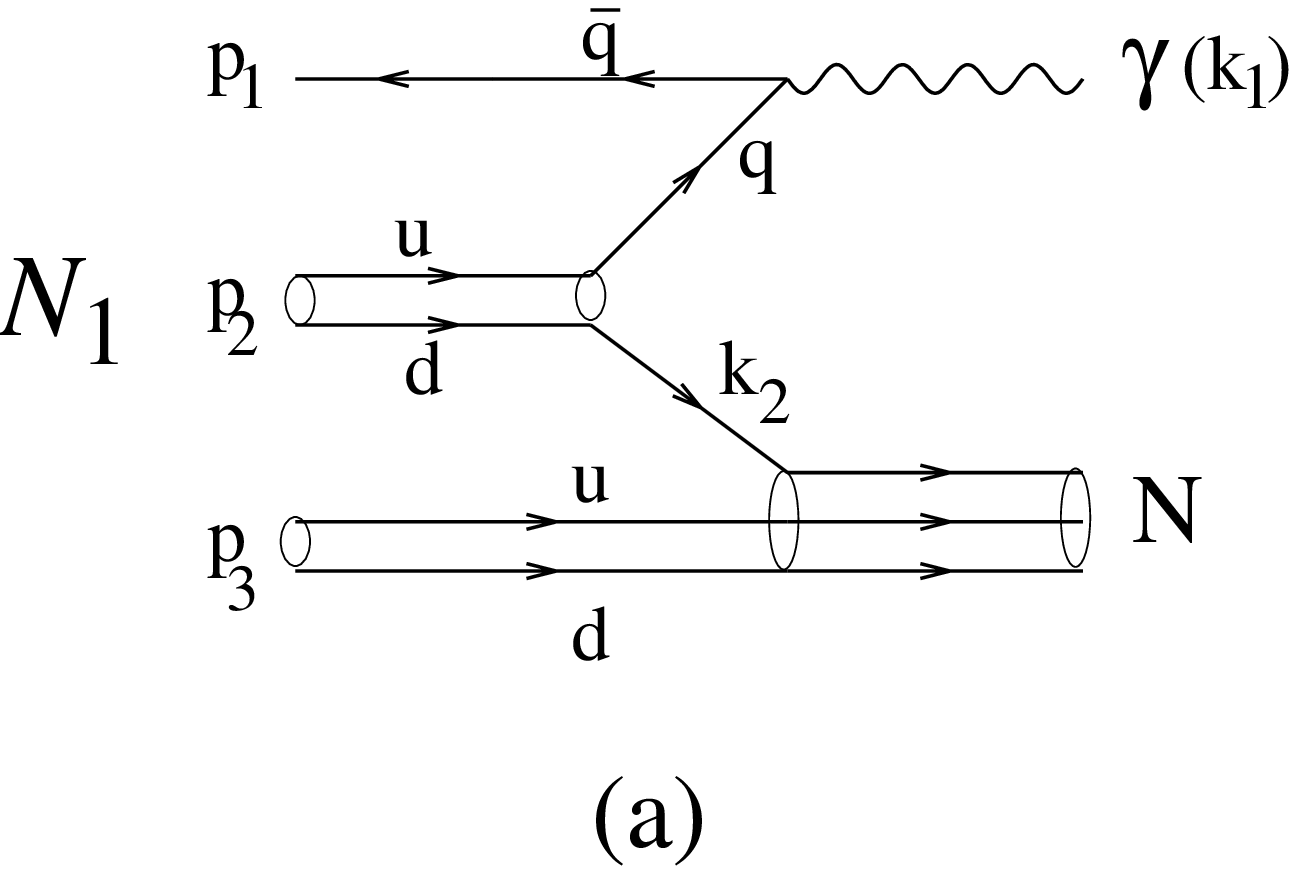}
\hskip 1in {\epsfxsize=2in \epsffile{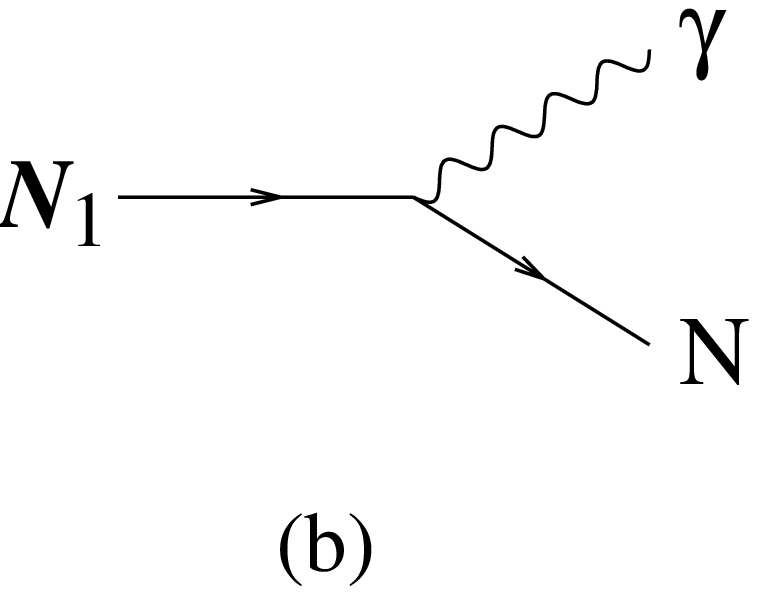}}}}
\caption{The radiative decay of the Roper N(1440)}
 \label{fig2}
\end{figure}
The leading interaction for the radiative decay of the
pentaquark octet in the chiral perturbation theory (Fig.~\ref{fig2}.~b)
is given as
\begin{equation}
{\cal L}_{{\cal O}B\gamma}^{(1)}= e_{*}\bar B\!\not\!\! A\,Q_{e} {\cal O}+{\rm h. c.}
\end{equation}
where $B$ is the normal baryon octet,
$A_{\mu}$ is the photon field, and  $Q_e$ is the electric charge matrix
acting on the anti-quark bound in the pentaquark octet, ${\cal O}$.
The effective electric charge $e_*$ is proportional to
the tunneling amplitude, $e^{-S_0({\cal N}_1)}$, since the
decay occurs by annihilation of a quark bound in a diquark after the other quark
in the diquark tunnels to another diquark.


For the radiative decay we get, after
summing over the photon polarization,
\begin{eqnarray}
\Gamma\left({\cal N}_1\to N\gamma\right)
=Q_e^2\frac{e_*^2}{2\pi}\,\frac{M_{{\cal N}_1}^2-m_N^2}{M_{{\cal N}_1}^3}
\times\left[2M_{{\cal N}_1}m_N-\left(M_{{\cal N}_1}-
m_N\right)^2\right]
=e_*^2\,Q_e^2\,156~{\rm MeV},
\end{eqnarray}
where $Q_e$ is the electric charge of the anti-quark in the pentaquark octet.
Since the decay amplitude is proportional to
the electric charge of the anti-quark we get at the leading order
$\Gamma({\cal N}_1\to n\gamma)=4\,
\Gamma({\cal N}_1\to p\gamma)$.

Comparing the decay processes ${\cal N}_1\to\,N\pi$ (Fig.~1.~a) and
${\cal N}_1\to\,N\gamma$ (Fig.~2.~a), we obtain
$e_*\,g_A=e\,g_{\pi\,N{\cal N}_1}$,
where $g_A\simeq0.75$ is the quark axial coupling in the quark model.
Using the coupling obtained from the decay width for $N(1440)\to\,N\pi$,
we get $e_*=0.4-0.44~e$. Then, the partial
radiative decay width becomes
$\Gamma\left({\cal N}_1\to p\gamma\right)=0.25\sim 0.31~{\rm MeV}$,
which is about 2 times larger that the estimate made by
PDG~\cite{foot1}.

Finally, we consider the three-body radiative decay of pentaquarks.
The diagram for $\Theta^+\to KN$ and the dominant
diagram~\cite{foot2} for $\Theta^+\to
KN\gamma$ in the diquark effective theory are shown in
Fig.~{\ref{fig3}}.
\begin{figure}[h]
\vskip 0.1in
\epsfxsize=2in
{\centerline{\epsffile{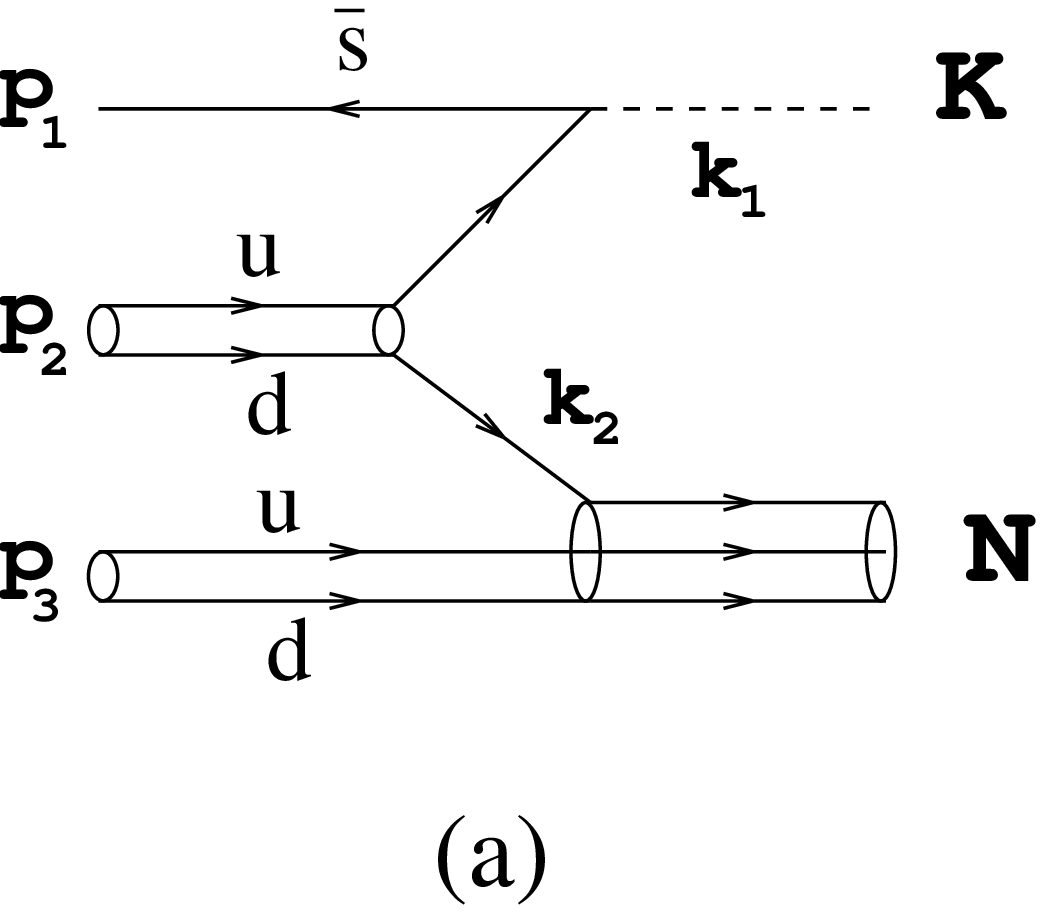}\hskip 1in {\epsfxsize=2in \epsffile{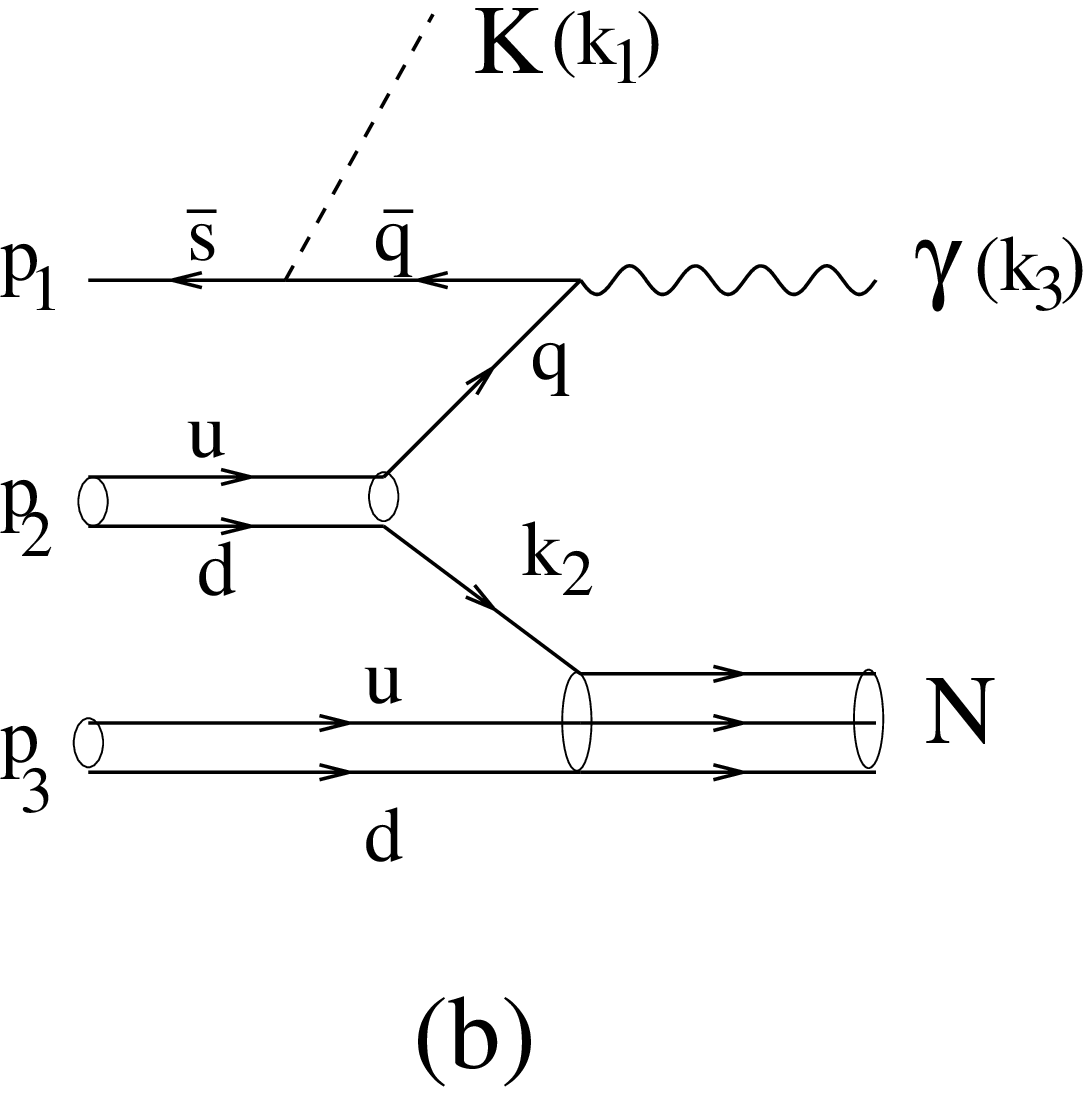}}}}
\caption{The decay modes of $\Theta^+$: (a) two-body decay and (b)
three-body radiative decay.}
 \label{fig3}
\end{figure}
The same processes  in the chiral perturbation theory are drawn in Fig.~{\ref{fig4}}.
\begin{figure}[h]
\vskip 0.1in
\epsfxsize=1.6in
{\centerline{\epsffile{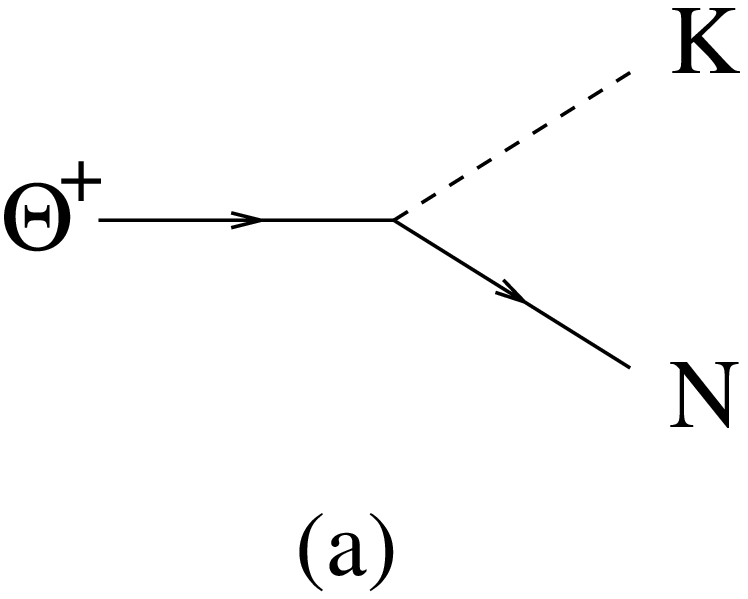}\hskip 1in
{\epsfxsize=2in \epsffile{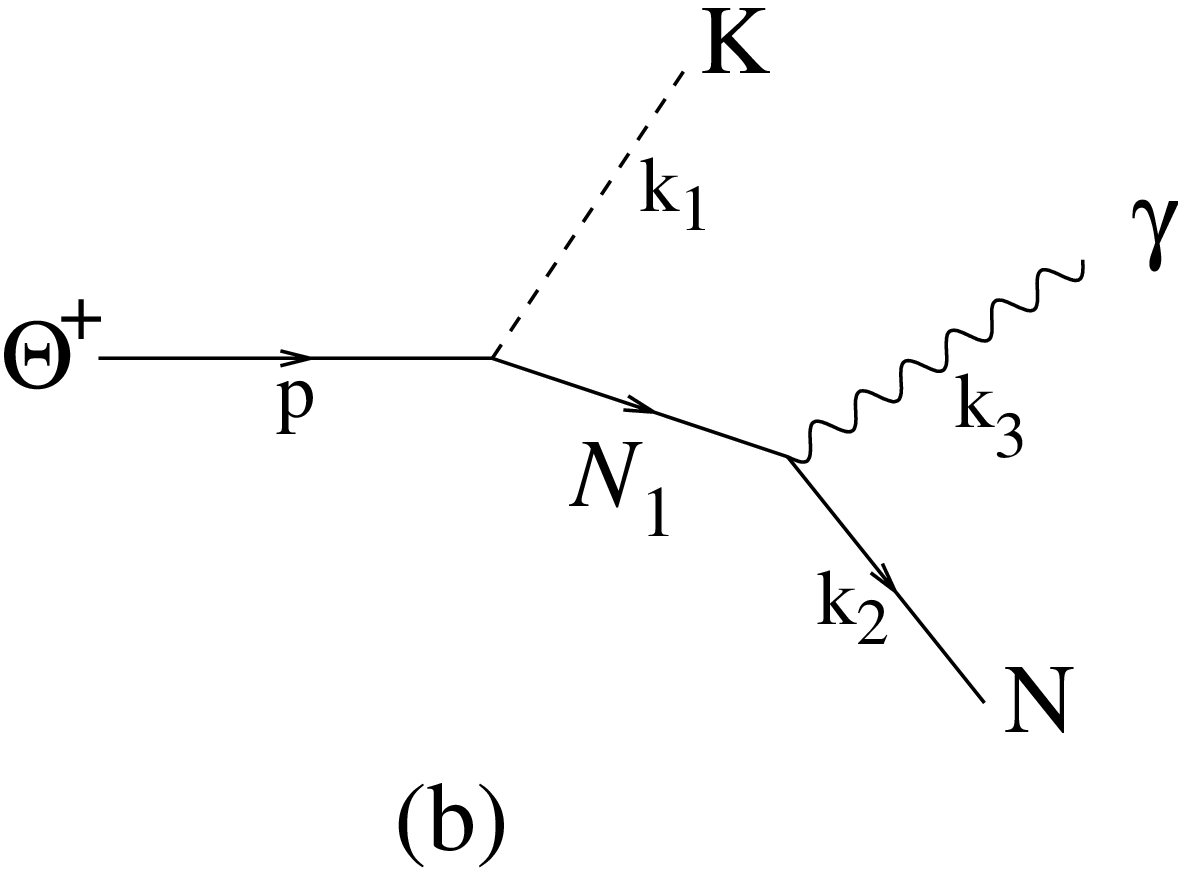}}}}
\caption{The decay processes of $\Theta^+$ in chiral perturbation theory}
 \label{fig4}
\end{figure}
The exotic baryons in the anti-decuplet are coupled to
the octet exotics by the meson octet, as well as  to the ground state baryon octet.
The leading interaction terms for the decay of the pentaquark anti-decuplet
and the pentaquark octet  are given  in the SU(3) limit as
\begin{eqnarray}
{\cal L}^{(1)}={\cal C}_{{\cal P}B}
\bar {\cal P} \gamma_5 \not\!\! a \, B
+g_{\cal PO}\bar {\cal P} \gamma_5 \not\!\! a \, {\cal O}
+{\cal D}_{{\cal O}B}\,{\rm Tr}\left(\bar{\cal O}\gamma_5\,\{\not\!\! a, B\}\right)
+{\cal F}_{{\cal O}B}\,{\rm Tr}\left(\bar{\cal O}\gamma_5\,[\not\!\! a, B]\right)
+{\rm h. c.}
\end{eqnarray}
where
${\cal P}$ and ${\cal O}$ are the anti-decuplet and octet pentaquarks, respectively,
and $a_{\mu}=i/2(\xi^{\dagger}\partial_{\mu}\xi- \xi\partial_{\mu}\xi^{\dagger})$
is the axial current of the meson octet, $\xi=\exp\,(i\pi_aT_a/f)$.
($T_a$'s are the SU(3) generators.)

The couplings ${\cal C_{{\cal P}B}}, {\cal D}_{{\cal O}B},
{\cal F}_{{\cal O}B}$ are proportional
to the axial coupling $g_A$, the
Yukawa coupling between the diquark and quarks, $g$,
and the tunneling amplitude $e^{-S_0}$~\cite{Hong:2004xn}.
The pentaquark octet will decay into the ground state baryon
octet and other particles.
But, only the radiative decay is allowed for the virtual octet
in the three-body decay of the anti-decuplet pentaquarks
because the allowed phase space is below the threshold
of any hadronic decays for the intermediate pentaquark octet.


The differential cross-section for the three-body decay
$\Theta^+\to K\,N\gamma$ (Fig.~{\ref{fig4}}.~b)  is given at
tree-level as
\begin{eqnarray}
\frac{{\rm d}\Gamma\left(\Theta^+\to K\,N\gamma\right)}
{{\rm d}|\vec k_3|\,{\rm d}\cos\theta_{\gamma}}
=\frac{g_{\Theta^+{\cal N}_1K}^2\,e_*^2\,Q_e^2}{128\pi^3 }
\frac{m_K^3}{M_{\Theta^+}\,f_K^2}
\times F(|\vec k_3|, \cos\theta_{\gamma}\,;\,M_{\Theta^+},\,M_{{\cal N}_1},\, m_N,\,m_K),
\label{three}
\end{eqnarray}
where $m_K$  and $f_K\simeq 114~{\rm MeV}$
are the mass and the decay constant of kaons, respectively.
The photon energy is $|\vec k_3|$ and
$\theta_{\gamma}$ is the angle between the kaon momentum and the photon momentum.
The function $F(|\vec k_3|,\cos\theta_{\gamma}\,{\rm ;} \,M_i)$
has a peak when two momenta are
opposite. (See Fig.~\ref{fig5}.)
\begin{figure}[h]
\vskip 0.1in
\epsfxsize=2in
{\centerline{\epsffile{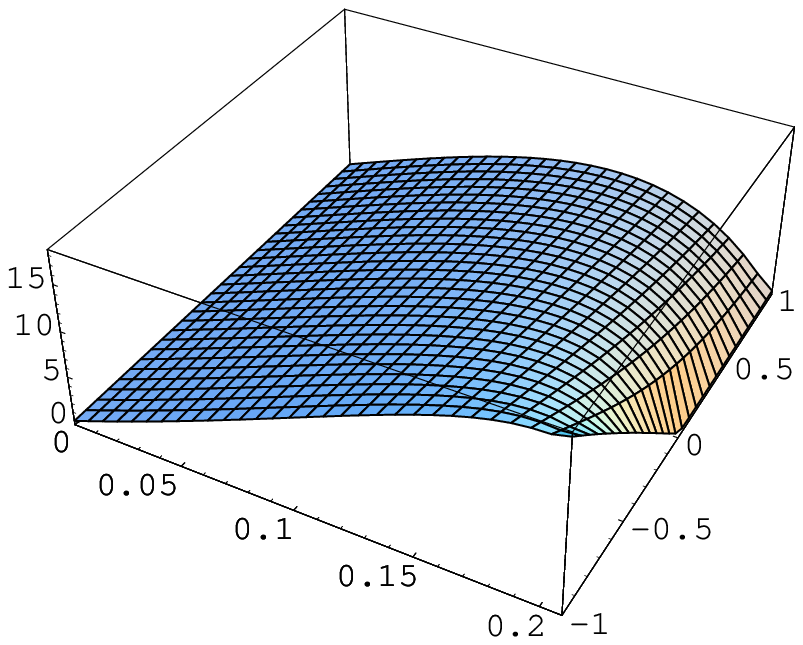}
\hskip 1in {\epsfxsize=2in \epsffile{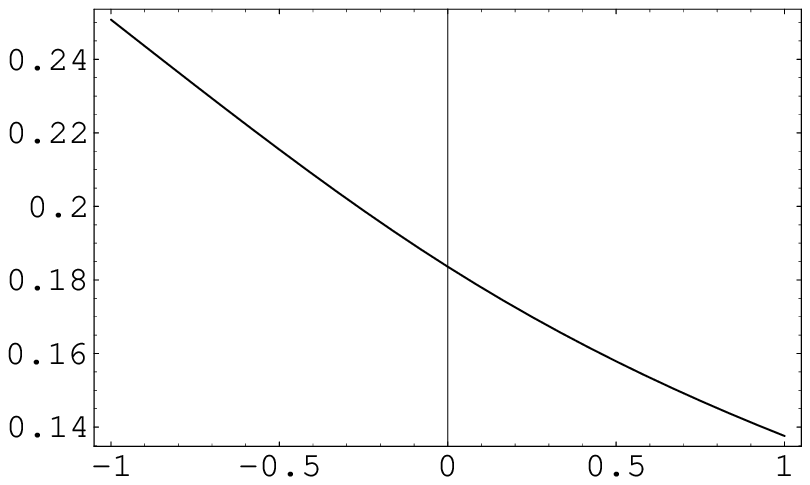}}}}
\caption{In the left the differential decay width for $\Theta^+\to\,K^+\,n\gamma$
is drawn
as a function of photon energy (in unit of $m_K$)
and $\cos\theta_{\gamma}$. In the right, ${\rm d}\Gamma/({\rm d}\cos\theta_{\gamma})$
is drawn in $0.1~{\rm MeV}$ as $\cos\theta_{\gamma}$ varies. }
 \label{fig5}
\end{figure}

Integrating over the angle $\theta_{\gamma}$, we get
the radiative decay width of the pentaquark $\Theta^+$
\begin{eqnarray}
\Gamma\left(\Theta^+\!\to\! K^+n\gamma \right)
=4\Gamma\left(\Theta^+\!\to\! K^0\,p\gamma \right)
\simeq 1.38\frac{e_*^2\,g_{\Theta^+{\cal N}_1K}^2}{256\pi^3}
\frac{m_K^4}{M_{\Theta^+}\,f_K^2}\simeq 0.034\sim 0.041\,{\rm MeV},
\end{eqnarray}
where we use $e_*=0.4\sim0.44\,e\,$
and $g_{\Theta^+{\cal N}_1K}\simeq2\sqrt{2}\,g_A=2.12$,
the naive quark model value.
In the SU(3) limit
the $\Theta^+$ coupling to the octet meson and the ground
state baryon is related to the physical couplings of $N(1440)$ and $N(1710)$,
which are mixed states of $\left|{\overline{10}}\right>$ and
$\left|8\right>$ :
${\cal C}_{{\cal P}B}=\sqrt{2}\,g_{\pi N\,{\cal N}_1}\sin\theta-
\sqrt{2}\,g_{\pi N\,{\cal N}_2}\cos\theta$.
Using the PDG values for the couplings (of the same sign)
and $\theta\simeq 20^{\circ}$, we find
$\left|{\cal C}_{{\cal P}B}\right|=0.031\sim 0.056$.
Since the two-body decay width of $\Theta^+$ in the
chiral perturbation theory~\cite{Ko:2003xx,Mehen:2004dy} is given as
$\Gamma(\Theta^+\to K\,N)={\cal C}_{{\cal P}B}^2\, 495~{\rm MeV}$,
we get $\Gamma(\Theta^+\to K^+\,n)=\Gamma(\Theta^+\to K^0\,p)=0.24\sim0.78~{\rm
MeV}$.
Therefore, our analysis shows that  the three-body
radiative decay contributes quite significantly
to the decay of $\Theta^+$: $4.6\sim 13.5\,\%$
in the $K^+\,n$ channel and $1.2\sim3.8\,\%$ in the $K^0\,p$ channel.
The slight violation of the isospin symmetry in the decay of $\Theta^+$ is
expected in the diquark picture.

We note that the systematic difference (of about $12~{\rm MeV}$)
in the measurement of the $\Theta^+$ mass
between $K^0\,p$ and $K^+\,n$, noted in~\cite{Rossi:2004rb},
may be due to the
missing photons in the radiative decay,
$\Theta^+\to K^0\,p\,\gamma$~\cite{foot3}.
In the $K^0\,p$ decay channel, the $\Theta^+$ mass is reconstructed from the
invariant mass, $M_{K^0p}\,$, of $K^0$ and $p$,
while it is measured from the missing mass in the $K^+n$ decay process.
But, when there is a missing photon
of momentum $\vec k_3$ in the $K^0\,p$ decay channel of $\Theta^+$,
the invariant mass is given in the rest frame of
$\Theta^+$ as
$M_{K^0p}^2=M_{\Theta^+}^2-2\,M_{\Theta^+}\,|\vec k_3|$.
The probability of finding the photon of energy in
$|\vec k_3|\sim |\vec k_3|+{\rm d}|\vec k_3|$ in the $K^0\,p$ channel is
\begin{equation}
\frac{1}{\Gamma_{\rm tot}}\,
\left(\frac{{\rm d}\Gamma_{{\rm 3body}}}{{\rm d}|\vec k_3|}\right)\,{\rm d}|\vec k_3|.
\end{equation}
where $\Gamma_{\rm tot}=\Gamma(\Theta^+\to K^0\,p)+\Gamma(\Theta^+\to K^0\,p\gamma)$
and ${\rm d}\Gamma_{\rm 3body}$ is the differential three-body radiative decay
width in Eq.~(\ref{three}), after integrating over $\theta_{\gamma}$.
Averaging over the photon energy distribution, we get
\begin{equation}
M_{K^0p}^2=M_{\Theta^+}^2-0.26\,M_{\Theta^+}\,m_K\,\frac{\Gamma_{\rm 3body}}{\Gamma_{\rm tot}}.
\end{equation}
Therefore, we find $M_{K^0p}\simeq1530 - 1528~{\rm MeV}$ for
$M_{\Theta^+}=1540~{\rm MeV}$ if
$\Gamma_{\rm 3body}\simeq0.15-0.19\,\Gamma_{\rm tot}$,
which is about $5-7$ times larger than our estimate on the radiative decay width.
The mass difference is due to missing photons, if the coupling
$g_{\Theta^+{\cal N}_1K}$ is about 3 times bigger than the naive quark model value
or ${\cal C}_{{\cal P}B}$ is about 1/3 of our estimate.


In conclusion,
we analyze the decay modes of the Roper ${ N}(1440)$ and
${N}(1710)$ in the Jaffe-Wilczek diquark model, provided that
they are the mixed states of the pentaquark octet and anti-decuplet.
We find that the experimental data on the partial decay widths
is consistent with the diquark model,
as the phenomenological couplings for the decay modes are proportional to
the tunneling amplitude of the diquark barrier.
We then predict the radiative decay of ${N}(1440)$
to be
$\Gamma_{10}({N}\to p\gamma)=\Gamma_{12}({N}\to n\gamma)/4
=0.25-0.31 ~{\rm MeV}$. The ratio is the unique prediction of
the pentaquark feature of $N(1440)$ in the diquark model.

Finally, the three-body radiative decay of pentaquarks
is shown to be quite enhanced in the Jaffe-Wilczek diquark model
because of the near degeneracy of the pentaquark anti-decuplet and octet and
the larger tunneling amplitude for the octet.
From the experimental data on ${N}(1440)$ and $N(1710)$,
we calculate the radiative decay width of $\Theta^+$ in chiral
perturbation theory to find about $0.04~{\rm MeV}$ in $K^+\,n\gamma$ channel
and $0.01~{\rm MeV}$ in $K^0\,p\gamma$ channel.
It is also shown that the mass difference of $\Theta^+$ in the
$K^0p$ and $K^+n$ decay channels may be accounted for by the missing photons
in the radiative decay.

The three-body radiative decay is quite significant
in the decay of $\Theta^+$, especially in the forward decay, since it has a peak
when the outgoing kaon and nucleon are collinear.
Our analysis suggests that one should take into account the
radiative decay in search of the pentaquark decay.





\acknowledgments We are grateful to V.~D. Burkert, M. Forbes, K. Hicks,
R. Jaffe, E. Lomon, V. Mohta,  D. Pirjol, M. Prasza{\l}owicz, M. Savage, I. Stewart, and
F. Wilczek for useful discussions. This work  is supported by
Korea Research Foundation Grant (KRF-2003-041-C00073) and also in
part by funds provided by the U.S. Department of Energy (D.O.E.)
under cooperative research agreement \#DF-FC02-94ER40818.


\begin{references}
\bibitem{Nakano:2003qx}
T.~Nakano {\it et al.}  [LEPS Collaboration],
Phys.\ Rev.\ Lett.\  {\bf 91} (2003) 012002
[arXiv:hep-ex/0301020].


\bibitem{Diakonov:1997mm}
D.~Diakonov, V.~Petrov and M.~V.~Polyakov,
Z.\ Phys.\ A {\bf 359}, 305 (1997)
[arXiv:hep-ph/9703373];
M. Praszalowicz, in Proceedings,
``Skyrmions And Anomalies",
M.~Jezabek and M.~Praszalowicz, eds., World Scientific (1987), p. 112;
A.~V.~Manohar,
Nucl.\ Phys.\ B {\bf 248}, 19 (1984);
%
M.~Chemtob,
Nucl.\ Phys.\ B {\bf 256}, 600 (1985).



\bibitem{Jaffe:2004ph}
For recent reviews, see R.~L.~Jaffe,
arXiv:hep-ph/0409065;
M.~Karliner and H.~J.~Lipkin,
arXiv:hep-ph/0307243;
M.~Praszalowicz,
arXiv:hep-ph/0410241.






\bibitem{trilling}
For a recent overview on the pentaquarks experiments,
see G. Trilling in~{\cite{Eidelman:2004wy}}.


\bibitem{Jaffe:2003sg}
R.~L.~Jaffe and F.~Wilczek,
Phys.\ Rev.\ Lett.\  {\bf 91} (2003) 232003
[arXiv:hep-ph/0307341].


\bibitem{Alt:2003vb}
C.~Alt {\it et al.}  [NA49 Collaboration],
Phys.\ Rev.\ Lett.\  {\bf 92}, 042003 (2004)
[arXiv:hep-ex/0310014].



\bibitem{Hong:2004xn}
D.~K.~Hong, Y.~J.~Sohn and I.~Zahed,
Phys.\ Lett.\ B {\bf 596}, 191 (2004)
[arXiv:hep-ph/0403205].


\bibitem{Cohen:2004gu}
T.~D.~Cohen,
Phys.\ Rev.\ D {\bf 70}, 074023 (2004) [arXiv:hep-ph/0402056].

\bibitem{Goeke:2004ht}
K.~Goeke, H.~C.~Kim, M.~Praszalowicz and G.~S.~Yang,
arXiv:hep-ph/0411195.


\bibitem{Mohta:2004xg}
V.~Mohta,
arXiv:hep-ph/0411247.


\bibitem{Eidelman:2004wy}
S.~Eidelman {\it et al.}  [Particle Data Group Collaboration],
Phys.\ Lett.\ B {\bf 592}, 1 (2004).

\bibitem{Hong:2004ux}
D.~K.~Hong and C.~j.~Song,
arXiv:hep-ph/0407274.

\bibitem{foot1}
In a recent experiment at JLab,
they found strong longitudinal response to $N(1440)$
in $\gamma^*+ p\to N(1440)$~\cite{Aznauryan:2004jd}.


\bibitem{foot2}
There are other tree diagrams for the radiative decay,
but they are all suppressed, since they do not have ${\cal N}_1$ or
have more powers of the electric charge.

\bibitem{Mehen:2004dy}
T.~Mehen and C.~Schat,
Phys.\ Lett.\ B {\bf 588}, 67 (2004)
[arXiv:hep-ph/0401107].


\bibitem{Ko:2003xx}
P.~Ko, J.~Lee, T.~Lee and J.~h.~Park,
arXiv:hep-ph/0312147;









\bibitem{Rossi:2004rb}
P.~Rossi  [CLAS Collaboration],
arXiv:hep-ex/0409057;
K.~Hicks,
arXiv:hep-ex/0412048.



\bibitem{foot3}
The author thanks K. Hicks for discussions on this.


\bibitem{Aznauryan:2004jd}
I.~G.~Aznauryan, V.~D.~Burkert, H.~Egiyan, K.~Joo, R.~Minehart and L.~C.~Smith,
arXiv:nucl-th/0407021.





\end{references}
\end{document}